\documentclass{svproc}
\usepackage{graphicx}
\usepackage{amsmath}
\usepackage{url}
\usepackage{tikz}
\usetikzlibrary{automata, calc, positioning,arrows.meta,shapes, arrows, intersections}
\usepackage{pgfplots}
\usepackage{pgfplotstable}
\pgfplotsset{compat=1.18}
\usepackage{xcolor}
\usepackage{array}

\AtBeginDocument{%
  }

\begin{document}

\title{Llms, Virtual Users, and Bias: Predicting Any Survey Question Without Human Data}
\titlerunning{LLMs and prediction of survey questions}

\author{Enzo Sinacola$^{1,2}$ \and Arnault Pachot$^{2}$ \and Thierry Petit$^{2,3}$}
\authorrunning{Sinacola et al.}

\institute{Emotia, 27 rue Marbeuf, 75008 Paris, France\\
}
\maketitle

\begin{center}
\email{$^1$enzosin.1005@gmail.com} \\
\email{$^{1,2,3}$\{contact,apachot,tpetit\}@emotia.com} \\
\email{$^3$tpetit19@gmail.com}
\end{center}

\begin{abstract}
  Large Language Models (LLMs) offer a promising alternative to traditional survey methods, potentially enhancing efficiency and reducing costs. In this study, we use LLMs to create virtual populations that answer survey questions, enabling us to predict outcomes comparable to human responses. We evaluate several LLMs—including GPT-4o, GPT-3.5, Claude 3.5-Sonnet, and versions of the Llama and Mistral models—comparing their performance to that of a traditional Random Forests algorithm using demographic data from the World Values Survey (WVS). LLMs demonstrate competitive performance overall, with the significant advantage of requiring no additional training data. However, they exhibit biases when predicting responses for certain religious and population groups, underperforming in these areas. On the other hand, Random Forests demonstrate stronger performance than LLMs when trained with sufficient data. We observe that removing censorship mechanisms from LLMs significantly improves predictive accuracy, particularly for underrepresented demographic segments where censored models struggle. These findings highlight the importance of addressing biases and reconsidering censorship approaches in LLMs to enhance their reliability and fairness in public opinion research.
\end{abstract}

\section{Introduction}
Traditional survey methods, while effective, are often time-consuming and costly, requiring extensive resources to survey real individuals and validate responses. The advent of Large Language Models (LLMs) offers a promising alternative, potentially enhancing efficiency, reducing costs, and minimizing biases in survey research. Applications include automated survey design, data collection, and context-based analysis~\cite{mellon2019survey}. 
Recent studies have highlighted the potential of LLMs to forecast survey responses and simulate public reactions \cite{santurkar_whose_2023}. 

Our study examines the efficacy of LLMs in public opinion polling compared to traditional machine learning approaches such as Random Forests, through the creation of virtual populations that answer survey questions. Specifically, we assess the accuracy of LLMs in predicting survey responses across various demographic groups and evaluate the impact of bias mitigation techniques in LLMs.
Our results demonstrate that LLMs are competitive with traditional models like Random Forests. A key advantage is that they require no additional training data. However, they face challenges in accurately predicting responses for diverse or underrepresented populations. In this context, we show that removing censorship consistently improves predictive accuracy, especially in demographic groups where censored models struggle. These findings underscore the importance of addressing biases in LLMs to enhance their reliability in public opinion research.

\section{Previous Work}

Recent studies have explored the use of Large Language Models (LLMs) in public opinion polling. Argyle et al.~\cite{argyle_out_2022} demonstrated that LLMs like GPT-3 could replicate sub-populations by leveraging socio-demographic data, achieving high correlation with actual survey responses, especially in politically affiliated groups. However, their work also revealed biases, particularly regarding race and gender. 
To address these biases, Santurkar et al.\cite{santurkar_whose_2023} introduced the OpinionsQA dataset, revealing significant discrepancies between LLM-generated opinions and actual demographic views, stressing the need for better alignment in LLM predictions. Similarly, Durmus et al.\cite{durmus_towards_2024} showed that LLMs often struggle to represent global perspectives accurately, favoring responses from Western populations.
Although approaches such as fine-tuning LLMs with cross-sectional survey data have aimed to improve predictive accuracy~\cite{kim_ai-augmented_2024}, challenges persist in achieving reliability across diverse contexts. Bisbee et al.\cite{bisbee_synthetic_2023} warned of significant biases when using synthetic survey data from LLMs, citing issues like overconfidence in predictions and prompt sensitivity. 
These studies highlight the need for further research to address biases and improve model alignment with real-world data. Our study builds on this body of work by comparing LLMs with traditional models and evaluating the impact of censorship mechanisms on predictive accuracy across diverse demographic groups.

\section{Methodology}
\subsection{Dataset Selection}
For this study, we utilize the World Values Survey (WVS) Wave 7 \cite{wvs2020}, a comprehensive dataset collected between 2017 and 2022, which surveys 94,278 individuals from 64 countries. The dataset was selected for its rich diversity of demographic information, including age, religion, education, and opinions on societal issues such as politics, economics, and culture. This diversity allows for robust comparisons across different demographic subgroups. Additionally, WVS is widely recognized in the social sciences for its methodological rigor and comprehensive scope.
The key factors for choosing this dataset are as follows: \textbf{Diverse population representation}, covering various regions, cultures, and social backgrounds; \textbf{Comprehensive opinion data}, facilitating comparisons across numerous demographic characteristics; \textbf{Large sample size}, allowing for robust statistical analysis across multiple subgroups; and \textbf{High scientific recognition and reliability}, as this dataset is widely cited in social science research, enhancing the credibility of our analysis.

\subsection{Methodology Overview}
The methodology of this study is structured into four key phases, each designed to address specific research objectives outlined in the introduction.
\paragraph{Phase 1: Prompt and Temperature Optimization for LLM} This phase serves as a foundational step to identify the optimal prompt and temperature settings. For this first step, we only use Mistral-7B~\cite{Mistral}, chosen for its balance between performance and resource efficiency. This phase aims to determine which configurations yield the most accurate predictions on survey responses.

\paragraph{Phase 2: Comparative Testing of Multiple Models} In this phase, we broaden our evaluation by comparing the performance of several LLMs and a traditional machine learning algorithm, Random Forests, on binary questions. We use Random Forests models due to their robustness. They are rarely subject to over-fitting when dealing with outliers, allowing them to generalize quite well. Unlike LLMs or neural networks in general, they require much fewer computational resources and can be trained very fast, which makes them a cost-effective and competitive alternative for comparison. They are trained one per question and on different portions of the dataset, to evaluate their performance based on the quantity of data in training.

The models are tested on a consistent sample of 384 individuals from the survey data.\footnote{The sample size of 384 individuals per group was determined based on the desire to achieve a confidence level of 95\% with a margin of error of 5\%, which is standard in social science research for statistical significance.} The results are compared to assess which models provide the most accurate predictions under uniform conditions using the optimal prompt and temperature settings identified in the previous phase.
The models tested include GPT-4o and GPT-3.5 \cite{GPT}, Claude 3.5-Sonnet \cite{Claude3.5Sonnet}, as well as open-source models Llama3-8B, Llama3.1-8B \cite{Llama}, and Mistral-7B \cite{Mistral}, along with their Dolphin-enhanced versions\footnote{Dolphin-enhanced versions are uncensored, with minimal bias correction. We hypothesized that this feature might affect result quality, an assumption confirmed by our findings.} \cite{DolphinLlama3} \cite{DolphinLlama3.1}\cite{DolphinMistral}, and Random Forests models.
By including Random Forests in this comparative testing phase, we aim to assess whether traditional machine learning models can rival or exceed the predictive capabilities of LLMs. This comparison will help determine whether the complexity of LLMs is necessary for accurate prediction in this context or whether simpler models like Random Forests are sufficient.
\paragraph{Phase 3: Performance Across Ethnic and Religious Groups} The third phase investigates how LLMs perform across different population segments on binary questions, specifically ethnic and religious groups. By dividing the dataset into these categories, this phase evaluates whether the LLMs exhibit any biases or performance disparities when predicting survey responses for different demographic groups. The analysis focuses on several major ethnic and religious groups to assess the robustness and fairness of the models when handling diverse populations.
The models tested are Llama3-8B, Llama3.1-8B, Dolphin-Llama3-8B, Dolphin-Llama3.1-8B, and Mistral-7B.
\paragraph{Phase 4: Censored vs. Uncensored LLM Comparison} Building on the findings from the third phase, this phase compares the performance of censored and uncensored versions of the LLMs across the same demographic groups. The goal is to determine whether the censorship mechanisms in the original models hinder their ability to generate accurate predictions, particularly for more nuanced or sensitive demographic questions. 

The models tested are Llama3-8B and Llama3.1-8B, both with their Dolphin version.

\subsection{Features and Survey Questions}
We distinct between demographic characteristics and opinions in the data extracted from the World Values Survey (WVS) dataset: demographic characteristics can be collected directly or through questions, allowing for simulation in virtual populations, whereas opinions must be gathered through surveys, reflecting individuals' subjective responses. 

\begin{enumerate}
    \item \textbf{Demographic Characteristics (Input Features)}: These features represent the demographic profile of the surveyed individuals, such as: 
Age, Gender, Education level, Religion, Marital status, Socioeconomic status, Country of origin. These features amount to a total of 35 input features..
     
   These demographic characteristics serve as input to both LLMs and Random Forests, allowing them to predict respondents' opinions.

    \item \textbf{Predicted answers (Target Variables)}: These target variables consist of 196 questions and are the survey responses that the models attempt to predict based on the demographic inputs. The target variables are split into three groups:
        \begin{itemize}
            \item \textbf{Questions with Scales or Ranges}, such as Likert-scale questions where respondents express their agreement or disagreement with a statement on a scale of 1 to 5. 146 questions were classified as such.
            \item \textbf{Questions without Scales or Ranges}, including categorical responses with no inherent ordering, such as multiple-choice questions on political preferences. We counted 50 of them.
            \item \textbf{Binary Questions}: Simple yes/no questions, such as whether respondents approve or disapprove of a certain policy. Questions that only refer to opinion, or thoughts. There are 30 of them in the Dataset. These are the questions we focused on for the majority of our tests. Since these questions are a minority in the Dataset, generating the answers only on these questions allowed us to considerably reduce the calculation times.
        \end{itemize}
\end{enumerate}

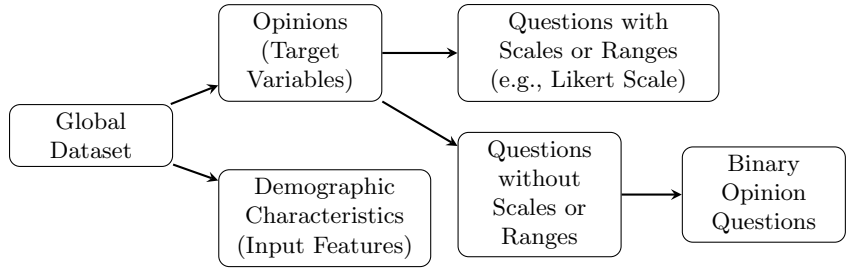
\begin{figure}[h!]
\centering
\begin{tikzpicture}[node distance=1cm, auto]
\footnotesize

    \tikzstyle{block} = [rectangle, draw, text width=6em, text centered, rounded corners, minimum height=2em]
    \tikzstyle{arrow} = [thick,->,>=stealth]
    
    \node [block] (dataset) {Global\\ Dataset};

    \node [block, text width=8em, below right=0.03cm and 0.6cm of dataset] (input) {Demographic\\ Characteristics\\ (Input Features)};

    \node [block, above right=0.03cm and 0.6cm of dataset] (target) {Opinions (Target Variables)};

    \draw [arrow] (dataset) -- (input);
    \draw [arrow] (dataset) -- (target);

    \node [block, text width=10em, right=1cm of target] (ellipse1) {Questions with Scales or Ranges (e.g., Likert Scale)};
    \node [block, below right=0.4cm and 1cm of target] (ellipse2) {Questions without\\ Scales or Ranges};
    \node [block, right=0.8cm of ellipse2] (ellipse3) {Binary Opinion\\ Questions};

    \draw [arrow] (target) -- (ellipse1);
    \draw [arrow] (target) -- (ellipse2);
    \draw [arrow] (ellipse2) -- (ellipse3);

\end{tikzpicture}
\caption{Dataset division into input features and target variables}
\label{fig:dataset_division}
\end{figure}

This division of the dataset into input features and target variables is critical for assessing how well the models can predict answers from demographic data. Figure \ref{fig:dataset_division} illustrates this division.

\subsection{Population Sampling and Generation}
We focused on two key demographic characteristics to guide the LLM in predicting survey responses: religion and ethnicity. These choices are motivated by their significant impact on individual responses and their relevance in studying biases and cultural differences in predictions.
The dataset provides specific ethnic labels for each country, which we used directly in our analysis. However, due to the wide variety of populations, many ethnicities were too specific or underrepresented for meaningful analysis. To ensure robustness, we grouped them  into broader categories that are geographically and culturally coherent, ensuring sufficient representation for each group.
\begin{table}[h!]
\centering
\scriptsize
\begin{tabular}{p{3cm}p{10cm}}
\hline
\textbf{Region}            & \textbf{Countries and Ethnicities}                                              \\ \hline
\textbf{North Africa}      & Morocco (Arab), Tunisia (Arab), Egypt (Arab), Libya (Arab)                      \\ \hline
\textbf{Sub-Saharan Africa}& Nigeria (Hausa), Ethiopia (Oromo), Kenya (Kikuyu), Zimbabwe (Ndebele)           \\ \hline
\textbf{Latin America}     & Brazil (Brown), Ecuador (Mestizo), Mexico (Light brown), Argentina (Light brown) \\ \hline
\textbf{North America}     & Canada (Caucasian white), USA (White, non-Hispanic)                             \\ \hline
\textbf{Europe}            & Netherlands (Caucasian white), Romania (Caucasian white), UK (British), Greece (Caucasian white) \\ \hline
\textbf{East Asia}         & China (Chinese), South Korea (East Asian), Hong Kong (Chinese), Mongolia (Khalkh) \\ \hline
\textbf{Southeast Asia}    & Indonesia (Javanese), Thailand (Thai), Vietnam (Kinh), Malaysia (Malay)         \\ \hline
\textbf{South Asia}        & Pakistan (Punjabi), Bangladesh (Bengali)                                        \\ \hline
\textbf{Middle East}       & Iran (Persian), Iraq (Arab)                                                     \\ \hline
\end{tabular}
\end{table}

By randomly selecting an equal number of individuals for each ethnicity within these groups, we maintained balanced representation and avoided biases that could arise from unequal sample sizes. This approach also enhances computational efficiency while ensuring the statistical validity of our comparisons.

For the religious analysis, we selected eight major religious groups, each represented by a sample of 384 individuals. 
These groups are \textbf{Islam}, \textbf{Buddhism}, \textbf{Hinduism}, \textbf{Protestantism}, \textbf{Eastern} \textbf{Orthodoxy}, \textbf{Evangelicalism}, \textbf{Non}\textbf{-religious individuals}, and \textbf{Roman} \textbf{Catholicism}.

\subsection{Objectives}
The main objectives of this study are: 
\begin{enumerate} 
\item To compare the performance of various LLMs and a traditional machine learning algorithm (Random Forests) in predicting survey responses based on demographic data.
\item To assess the accuracy of the models across different demographic groups, particularly ethnic and religious groups, and identify any inherent biases. 
\item To evaluate the impact of censorship mechanisms on LLM performance, determining whether they hinder the models' ability to generate accurate predictions, especially for sensitive demographic questions. 
\end{enumerate} 
The three objectives allow us to address key questions in the field of predictive modeling, including the ethical implications of using LLMs and their ability to handle sensitive or diverse demographic data effectively.

\section{Results}
This section presents the analysis results on temperature and prompt variations, LLM and Random Forest performance, average model accuracy across ethnic and religious groups, and comparisons of censored versus uncensored models.

\subsection{Comparative Testing}
The Mistral-7B model was tested extensively to evaluate the impact of varying prompt structures and temperature settings on survey prediction accuracy. Prompt structures shape how the model interprets and generates responses, while temperature settings control the randomness and diversity of the output. Our goal was to understand how these factors influence the model's ability to predict survey responses accurately and consistently. 

\paragraph{Prompt Influence}
Our tests indicate that prompt variations have a modest but noticeable impact on the model's performance across different question types. When evaluating all survey questions, accuracy varied slightly with different prompts, ranging from 36.7\% to 39.3\%. For categorical questions, prompts influenced average accuracies between 49.9\% and 51.7\%, showing a small performance difference due to prompt wording. In binary question evaluations, the best prompt achieved an average accuracy of 67.11\%, with others close behind. These results demonstrate that while prompt design does affect performance across all, categorical, and binary questions, the influence is moderate and not critical to overall accuracy.
\paragraph{Temperature Sensitivity and Stability Tests}
Our findings show that temperature settings modestly affect model accuracy, with higher temperatures introducing instability and slight declines in performance. For categorical questions, accuracy for a single prompt varies from 51.14\% at 0.1 in temperature to 49.46\% at 1 in temperature. Another prompt showed a decrease from 50.80\% at moderate temperatures to around 50.13\% at high temperatures. In binary questions, accuracy varied slightly from 67.32\% at moderate temperatures to 66.98\% at the lowest temperature. 
Building on these findings, stability tests conducted over multiple runs using the same prompts and questions confirmed that lower temperature settings (0 and 0.001) produced the most consistent results across repeated runs, maintaining an average accuracy of 68.26\% for each run. Higher temperatures introduced slight variability, with fluctuations in accuracy across runs, highlighting the importance of using lower settings for reproducible results.

Both prompt structure and temperature settings must therefore be considered to optimize Mistral-7B's performance, with prompt variations having a more significant impact on accuracy than temperature adjustments. 

\subsection{Model Comparison}
In the second part of the study, we evaluated the performance of all our LLMs and Random Forests models. The LLMs were tasked with predicting the answers of 384 individuals to the 30 binary questions from the dataset. The Random Forests models were trained using varying portions of the dataset, excluding the 384 individuals, and tested on these 384 to ensure a fair comparison with the LLMs.
The Random Forests models demonstrated strong performance even when trained on small subsets of the data. With only 1\% of the dataset, they achieved an average accuracy of 71.58\%, comparable to the top-performing LLMs. When the training dataset size was increased to 5\%, the accuracy improved to 73.40\%. When trained on 95\% of the dataset, the Random Forests models achieved their highest average accuracy of 74.93\%, outperforming all the LLMs.

\begin{figure}[h!] 
\centering 
\includegraphics[width=0.5\linewidth]{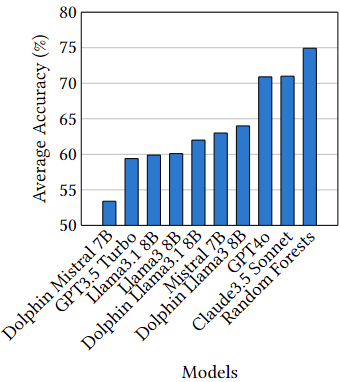} 
\caption{Average Accuracy by Models} 
\label{fig:accuracy_384_RF} 
\end{figure}

As shown in Figure \ref{fig:accuracy_384_RF}, among the LLMs, Claude 3.5 Sonnet and GPT-4o were the top performers, both reaching an accuracy of 71\%. Despite being outperformed by the Random Forests models, the LLMs demonstrated competitive results. Other models like Dolphin Llama3 8B and Mistral 7B achieved accuracies of 64\% and 63\%, respectively, while Dolphin Llama3.1 8B and Llama3 8B followed with 62\% and 60\%. GPT-3.5 Turbo recorded 59\%, with Dolphin Mistral 7B having the lowest accuracy of 53\%.

This comparison highlights the strong performance of the Random Forests models even when trained on a small portion of the dataset and emphasizes the competitive performance of the LLMs in predicting binary outcomes across the same individual sample.
Importantly, despite the strong Random Forest results with small training sets, LLMs need no training set. They adapt to all questions without needing to be retrained unlike Random Forests. These advantages make LLMs highly practical for conducting surveys when human responses are unavailable.

\subsection{Ethnic and Religious Group Analysis}
\begin{figure}[h!]
    \centering
    \begin{minipage}{0.4\textwidth}
        \centering
        \includegraphics[width=\linewidth]{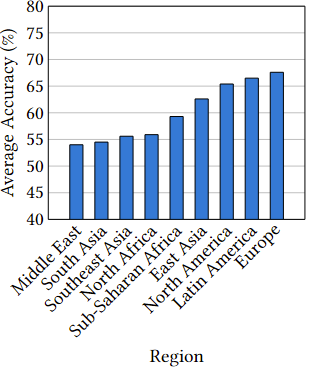}

        \caption{Average Accuracy by Region}
        \label{fig:ethnic_group_accuracy}
    \end{minipage}%
    \begin{minipage}{0.4\textwidth}
        \centering
        \includegraphics[width=\linewidth]{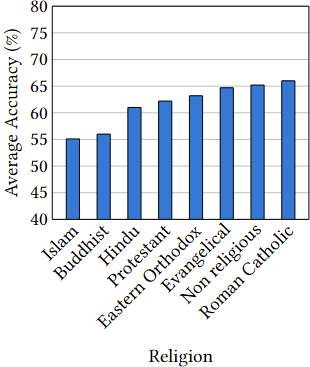}
        \caption{Average Accuracy by Religion}
        \label{fig:religion_group_accuracy}
    \end{minipage}
\end{figure}

Following the model comparison, we focused on a deeper performance analysis across demographic segments. The Ethnic and Religious Group Analysis section examines how the models perform specifically for different geographic and religious groups, with a consistent sample size of 384 individuals per group. This approach provides insights into model accuracy within culturally and regionally diverse populations, revealing strengths and limitations that may not be apparent in the broader model comparison.
The average accuracies are presented in Figures \ref{fig:ethnic_group_accuracy} and \ref{fig:religion_group_accuracy}.

\paragraph{Ethnic Group Performance :}
The results show varying accuracy across different groups. The Europe group achieved the highest accuracy, at 67.6\%, followed closely by Latin America at 66.5\%. The North America group scored 65.4\%, and East Asia scored 62.6\%. The Sub-Saharan Africa group achieved 59.3\%, while Southeast Asia and North Africa scored 55.6\% and 55.9\%, respectively. The Middle East and South Asia groups had the lowest scores, at 54\% and 54.5\%.

\paragraph{Religious Group Performance :}
The Roman Catholic group achieved the highest average accuracy, at 66.0\%, followed by the Non-religious group at 65.2\%. The Evangelical, Eastern Orthodox, and Protestant groups scored 64.7\%, 63.2\%, and 62.2\%, respectively. The Hindu group achieved an accuracy of 61.0\%, while the Buddhist and Islam groups scored 56.0\% and 55.1\%, respectively.

\subsection{Censored vs Uncensored Models Comparison}
This section compares censored and uncensored versions of specific LLMs, namely Llama3-8B and Llama3.1-8B. We aim to understand how moderation filters might influence predictive accuracy across religious and population groups, as previous evaluations demonstrated certain populations seeing notable drops in accuracy.
Unlike Random Forests, which are built on structured, manually selected features, LLMs may exhibit bias or censorship due to the nature of their training data and any applied moderation or filtering processes. 

\begin{figure}[h!]
    \centering
    \includegraphics[width=0.65\linewidth]{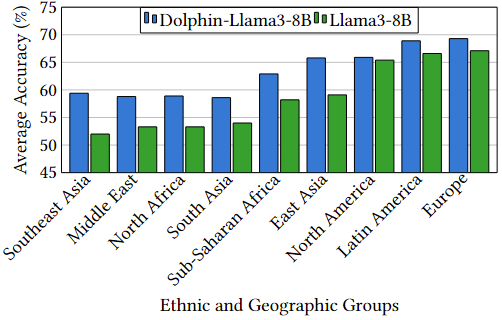}

\caption{Comparison of Model Accuracy Across Population Groups for Dolphin-Llama3-8B and Llama3-8B}
\label{fig:population_bias_comparison_accuracy}
\end{figure}

\paragraph{Population Group Accuracy Comparison:} Figure \ref{fig:population_bias_comparison_accuracy} compares the performance across population groups. Dolphin-Llama3-8B demonstrates a clear performance advantage, with the highest accuracy achieved in the Europe group at 69.3\%, followed closely by Latin America at 68.9\%. The lowest accuracy for this model is recorded in Southeast Asia at 59.4\%. For the censored Llama3-8B, accuracy is lower across all groups, with a maximum of 67.1\% in Europe and 52.0\% in Southeast Asia.

\begin{figure}[h!]
    \centering
    \includegraphics[width=0.65\linewidth]{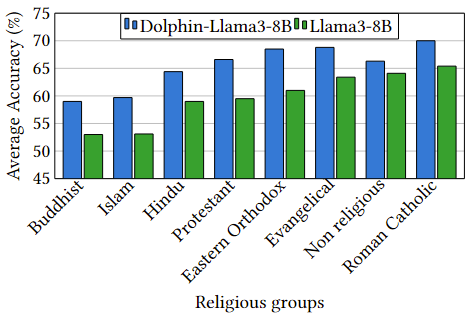}

\caption{Comparison of Model Accuracy Across Religions for Dolphin-Llama3-8B and Llama3-8B}
\label{fig:religion_bias_comparison_accuracy}
\end{figure}

\paragraph{Religious Group Accuracy Comparison:} Figure \ref{fig:religion_bias_comparison_accuracy} compares the average accuracy of Llama3-8B and Dolphin-Llama3-8B across religious groups. The results show that Dolphin-Llama3-8B consistently outperforms the censored model. The highest accuracy for Dolphin-Llama3-8B is observed in the Roman Catholic group at 70.0\%, while the Buddhist group records the lowest accuracy at 59.0\%. In contrast, Llama3-8B shows lower scores across all groups, with Roman Catholic at 65.4\% and Buddhist at 53.0\%.


\paragraph{Additional benchmarks:} The newer versions, Llama3.1-8B and Dolphin-Llama3.1-8B, show smaller performance gaps between the censored and uncensored models. On average, Dolphin-Llama3.1-8B outperforms the censored version by almost 2\%, reflecting a narrower gap in comparison to the 5\% difference observed with Llama3-8B.
We also looked at the performance of Mistral-7B on the different groups to confirm that the problem does not only come from Meta AI's Llama models. We found the same trend. Mistral performs better on the Europe, North America, Latin America regions. Regarding religions, the best are Christian religions and the non-religious group. 

\paragraph{Summary} Both figures highlight the consistent advantage of uncensored models, with Dolphin-Llama3-8B showing a more significant performance gap compared to Llama3.1-8B, which exhibits a smaller accuracy improvement between its censored and uncensored versions. This part also confirms that bias seems to affect a larger group of LLMs than just Meta AI models, particularly light models. 

\section{Discussion}
The primary aim of this study was to evaluate the efficacy of Large Language Models (LLMs) in predicting survey responses based on demographic data and to compare their performance with traditional machine learning models like Random Forests. Our results provide several insights into the capabilities and limitations of LLMs in the context of public opinion polling.
The key findings are the following:

\begin{itemize} 
    \item \textbf{LLM Performance vs. Random Forests:} LLMs such as GPT-4o and Claude 3.5-Sonnet demonstrated competitive performance, achieving very promising accuracy rates, around 71\%. However,  Random Forests models outperformed them with an accuracy of 74.93\% when trained on 95\% of the dataset.
    
    \item \textbf{Impact of Prompt Structure and Temperature:} Variations in prompt design modestly affected LLM performance, while temperature settings had a less significant effect, with lower temperatures yielding more consistent results.
    
    \item \textbf{Performance Across Demographic Groups:} LLMs showed varying accuracy across different ethnic and religious groups, performing better in European, Latin American, and Christian groups, and less accurately in South Asian, Middle Eastern, Islam, and Buddhist groups, indicating potential biases in the models.
    
    \item \textbf{Effect of Censorship Mechanisms:} Uncensored versions of LLMs consistently outperformed their censored counterparts, with significant improvements in predictive accuracy, especially in groups where censored models struggled.
\end{itemize}

These findings highlight important considerations. LLMs have demonstrated significant potential as effective tools for survey prediction, offering advantages such as scalability and the ability to handle complex linguistic patterns. These results can have a significant practical impact, as using LLMs allows for relatively accurate trend predictions for surveys in a very short time, without any prior dataset $\-$ just by considering demographic characteristics to generate populations. On the other hand, biases in training data may lead to underrepresentation of certain populations. Censorship mechanisms can hinder model performance. 

For future work, addressing the challenges of enhancing data diversity and refining censorship approaches is essential. By overcoming these issues, LLMs can become more reliable and accurate, boosting their effectiveness as tools for survey prediction. Additionally, while our study focused on smaller LLMs due to resource constraints, larger models could potentially deliver even better outcomes. Moving forward, our research will focus on fine-tuning, incorporating more diverse training data to reduce biases, and developing sophisticated censorship mechanisms that uphold ethical standards without compromising accuracy.

\section{Conclusion}
This study highlights the exciting potential of LLMs in predicting survey responses, demonstrating their capacity to simulate public opinion with flexibility and no need for training data. While Random Forest models trained on larger datasets achieved slightly higher accuracy, LLMs still showcased remarkable adaptability and scalability. Interestingly, variations in accuracy across demographic groups highlighted opportunities for fine-tuning, allowing us to identify areas where inclusivity can be enhanced. By further exploring the impact of content restrictions on LLM performance, we anticipate gaining valuable insights for optimizing virtual polling models.

\bibliographystyle{plain}
\bibliography{main}

\begin{thebibliography}{10}

\bibitem{Claude3.5Sonnet}
Anthropic.
\newblock Claude 3.5 sonnet, 2024.

\bibitem{argyle_out_2022}
Lisa~P. Argyle, Ethan~C. Busby, Nancy Fulda, Joshua Gubler, Christopher Rytting, and David Wingate.
\newblock Out of {One}, {Many}: {Using} {Language} {Models} to {Simulate} {Human} {Samples}, September 2022.
\newblock arXiv:2209.06899 [cs].

\bibitem{bisbee_synthetic_2023}
James Bisbee, Joshua Clinton, Cassy Dorff, Brenton Kenkel, and Jennifer Larson.
\newblock Synthetic {Replacements} for {Human} {Survey} {Data}? {The} {Perils} of {Large} {Language} {Models}, May 2023.

\bibitem{durmus_towards_2024}
Esin Durmus, Karina Nguyen, Thomas~I. Liao, Nicholas Schiefer, Amanda Askell, Anton Bakhtin, Carol Chen, Zac Hatfield-Dodds, Danny Hernandez, Nicholas Joseph, Liane Lovitt, Sam McCandlish, Orowa Sikder, Alex Tamkin, Janel Thamkul, Jared Kaplan, Jack Clark, and Deep Ganguli.
\newblock Towards {Measuring} the {Representation} of {Subjective} {Global} {Opinions} in {Language} {Models}, April 2024.
\newblock arXiv:2306.16388 [cs].

\bibitem{wvs2020}
Christian Haerpfer, Ronald Inglehart, Alejandro Moreno, Christian Welzel, Kseniya Kizilova, Jaime Diez-Medrano, Marta Lagos, Pippa Norris, Eduard Ponarin, and Bi~Puranen.
\newblock World values survey: Round seven - country-pooled datafile, 2020.

\bibitem{DolphinMistral}
Eric Hartford.
\newblock Dolphin mistral-7b, 2023.

\bibitem{DolphinLlama3}
Eric Hartford.
\newblock Dolphin llama3-8b, 2024.

\bibitem{DolphinLlama3.1}
Eric Hartford.
\newblock Dolphin llama3.1-8b, 2024.

\bibitem{kim_ai-augmented_2024}
Junsol Kim and Byungkyu Lee.
\newblock {AI}-{Augmented} {Surveys}: {Leveraging} {Large} {Language} {Models} and {Surveys} for {Opinion} {Prediction}, April 2024.
\newblock arXiv:2305.09620 [cs].

\bibitem{mellon2019survey}
Jonathan Mellon, Jack Bailey, Ralph Scott, James Breckwoldt, Marta Miori, and Phillip Schmedeman.
\newblock Do ais know what the most important issue is? using language models to code open-text social survey responses at scale.
\newblock {\em Research \& Politics}, 11(1), 2024.

\bibitem{Llama}
Meta-AI.
\newblock Llama family, 2024.
\newblock Llama3-8B(04/2024), Llama3.1-8B(07/2024).

\bibitem{Mistral}
Mistral-AI.
\newblock Mistral-7b, 2023.

\bibitem{GPT}
OpenAI.
\newblock Gpt, 2024.
\newblock Version gpt-4o-2024-05-13, gpt-3.5-turbo-0125.

\bibitem{santurkar_whose_2023}
Shibani Santurkar, Esin Durmus, Faisal Ladhak, Cinoo Lee, Percy Liang, and Tatsunori Hashimoto.
\newblock Whose {Opinions} {Do} {Language} {Models} {Reflect}?, March 2023.
\newblock arXiv:2303.17548 [cs].

\end{thebibliography}
\end{document}